\documentclass[aps,showpacs,nofootinbib,superscriptaddress]{revtex4}

\usepackage{amsmath}
\usepackage{amssymb}
\usepackage{graphicx}
\usepackage{float}
\usepackage{bm}
\usepackage{url}

\newcommand{\sumint}{\kern 0.2 em {\textstyle\sum} \kern -1.1 em \int_X}
\newcommand{\ii}{{\rm i}}

\begin{document}

\title{Single-spin asymmetry in dihadron production in SIDIS off the longitudinally polarized nucleon target}
\author{Wei Yang}
\affiliation{School of Physics, Southeast University, Nanjing 211189, China}
\author{Xiaoyu Wang}
\affiliation{School of Physics and Engineering, Zhengzhou University, Zhengzhou, Henan 450001, China}
\author{Yongliang Yang}
\affiliation{School of Physics, Southeast University, Nanjing 211189, China}
\author{Zhun Lu}
\email{zhunlu@seu.edu.cn}
\affiliation{School of Physics, Southeast University, Nanjing 211189, China}

\begin{abstract}
We study the single longitudinal-spin asymmetry of dihadron production in semi-inclusive deep inelastic scattering process. We consider the collinear picture in which the transverse momentum of the final-state hadron pair is integrated out, such that the $\sin \phi_R$ azimuthal asymmetry arises from the coupling $h_L\, H_{1}^{\sphericalangle}$ as well as the coupling $g_1 \,\widetilde{G}^{\sphericalangle}$. We calculate the unknown twist-3 dihadron fragmentation function $\widetilde{G}^{\sphericalangle}$ using a spectator model which is successful in describing the dihadron production in the unpolarized process.
Using the spectator model results for the quark distributions and dihadron fragmentation functions, we estimate the $\sin \phi_R$ asymmetry of dihadron production in SIDIS at the kinematics of COMPASS and compare it with the COMPASS preliminary data. In addition, the prediction on the $\sin \phi_R$ asymmetry at the typical kinematics of the future Electron Ion Collider is also presented.
In order to test the reliability of the spectator model estimate, we compare the model result for the distribution $h_L$ with the Wandzura-Wilczek approximation for that distribution, and compare $H_{1}^{\sphericalangle}$ with the existing parametrization.
Although the asymmetry is dominated by the $h_L H_{1}^{\sphericalangle}$ term, we find that the contribution from the $g_1\, \widetilde{G}^{\sphericalangle}$ term should also be taken into account in certain kinematical region.
\end{abstract}
\maketitle

\section{Introduction}

\label{Sec.introduce}

Understanding the partonic structure of the nucleon and the fragmentation mechanism of hadrons are the main tasks in QCD and hadronic physics.
The azimuthal asymmetries in semi-inclusive deep inelastic scattering (SIDIS) process have been recognized as useful tools for these quests.
The full description of SIDIS includes a set of parton distribution functions (PDFs) and fragmentation functions (FFs).~\cite{Kotzinian1995,9510301,Bacchetta2007}.
In recent years, the study of hadron pair production in SIDIS has received a lot of attention.
The dihadron FF (DiFF), which describes the probability that a quark fragmented into two hadrons: $q\to H_1 H_2 X$, appears in this process.

The unpolarized DiFFs were introduced in Ref.~\cite{Konishi1978}, and their evolution equations have been investigated in Refs.~\cite{Vendramin,Sukhatme,Majumder}.
The study of DiFFs was extended to the polarized cases in Refs.~\cite{hep-ph/9305309,hep-ph/9411444,hep-ph/9709322} in order to explore the transverse spin phenomena of the nucleon.
Particularly, the chiral-odd DiFF $H_1^{\sphericalangle}$~\cite{
Bianconi:1999cd,Radici:2001na,Bacchetta:2002ux} plays an important role in accessing transversity distribution, as it couples with $h_1$ at the leading-twist level in the collinear factorization.
In Refs.~\cite{Bacchetta:2011ip,Bacchetta:2012ty,Radici:2015mwa,Radici:2016lam,Radici:2018iag}, the authors applied this approach to extract $h_1$ from SIDIS and proton-proton collision data with the parameterized result for $H_1^{\sphericalangle}$~\cite{Courtoy:2012ry}.
Recently, there is also proposal~\cite{Matevosyan:2017liq,Matevosyan:2018icf} to probe the quark helicity through the helicity-dependent DiFF $G_1^\perp$ appearing in $e^+e^-$ annihilation.
Meanwhile, the calculations of the DiFF were carried out by the spectator model~\cite{hep-ph/9907488,Radici:2001na,Bacchetta:2006un,Bacchetta:2008wb} and by the Nambu-Jona-Lasinio (NJL) quark model~\cite{Matevosyan:2013aka,Matevosyan:2013eia,Matevosyan:2017uls}.

The leading-twist differential cross-section in polarized SIDIS, containing different azimuthal modulations involving dihadron fragmentation, was presented in Ref.~\cite{Radici:2001na}.
The study was then extended to the subleading twist in Ref.~\cite{Bacchetta:2003vn}, where the issue of gauge invariance of the DiFFs was also discussed.
In this formalism, the structure functions are expressed as the convolution of the distribution function and the DiFF.
Experimentally, hadron pair productions off both the unpolarized target and the transversely polarized target have been measured by the HERMES collaboration~\cite{Airapetian:2008sk} and the COMPASS collaboration~\cite{Adolph:2012nw,Adolph:2014fjw}.
Very recently, preliminary results on the azimuthal spin asymmetries in hadron pair production off the longitudinally polarized proton target were also obtained by the COMPASS collaboration~\cite{S.Sirtl}.
When the transverse momentum of the hadron pair is integrated out and the incident lepton beam is unpolarized, only one modulation -- the $\sin\phi_R$ azimuthal angle dependence -- remains.
Here $\phi_R$ is the angle between the lepton plan and the two-hadron plane.
The preliminary COMPASS measurement showed a clearly positive $\sin\phi_R$ asymmetry within experimental precision.
In the parton model, two sources~\cite{Bacchetta:2003vn} contribute to this asymmetry, one is the coupling of the twist-3 distribution $h_L$ and the twist-2 DiFF $H_1^{\sphericalangle}$, the other is the twist-3 DiFF $\tilde{G}^{\sphericalangle}$ combined with the helicity distribution $g_1$.

In this work, we study the $\sin\phi_R$ asymmetry by adopting the spectator model results for the distribution functions and fragmentation functions. We not only take into account the coupling $h_L \,H_1^{\sphericalangle}$, but also investigate the role of the T-odd DiFF $\tilde{G}^{\sphericalangle}$, which encodes the quark-gluon-quark correlation and has not been considered in previous studies.
It was suggested in Ref.~\cite{Metz:2012ct} that the fragmentation contribution in the twist-3 collinear framework may be also important for the SSA in $pp$ collision. Later phenomenological analysis~\cite{Kanazawa:2014dca} showed that, besides the contribution of the twist-3 collinear distribution functions, twist-3 fragmentation functions are also necessary for describing the SSA data in both SIDIS and $pp$ collision~\cite{Adams:2003fx,Abelev:2008af,Adamczyk:2012xd,Lee:2007zzh} in a consistent manner~\cite{Kang:2011hk}.
To calculate $\tilde{G}^{\sphericalangle}$, we apply a spectator model in Ref.~\cite{Bacchetta:2006un} where the parameters of the model is tuned to the output of the PYTHIA event generator.
Similar model has also been applied to calculate the single hadron fragmentation functions~\cite{Bacchetta:2007wc}.
we adopt the approach in Refs.~\cite{Lu:2015wja,1607.01638} in order to generate the gluon rescattering effect needed for nonzero T-odd fragmentation functions.
Using the model results for the distributions and DiFFs, we estimate the $\sin\phi_R$ asymmetry at COMPASS kinematics and compare it with the COMPASS preliminary data.

This paper is organized in the following way. In Sec.~\ref{Sec.formalism}, we review the theoretical framework of the $\sin\phi_R$ azimuthal asymmetry of dihadron production in unpolarized lepton beam scattered off a longitudinally polarized proton target. In Sec.~\ref{Sec.Model calculation}, we use a spectator model to calculate the twist-3 dihadron fragmentation function $\widetilde{G}^{\sphericalangle}$.
In Sec.~\ref{Sec.numerical}, We make the numerical estimate of the $\sin\phi_R$ azimuthal asymmetry at the kinematics of COMPASS as well as EIC. We summarize this work in Sec.~\ref{Sec.conclusion}.

\section{Formalism of the $\sin \phi_R$ asymmetry of dihadron production in SIDIS}

\label{Sec.formalism}

\begin{figure}
  \centering
  \includegraphics[width=0.48\columnwidth]{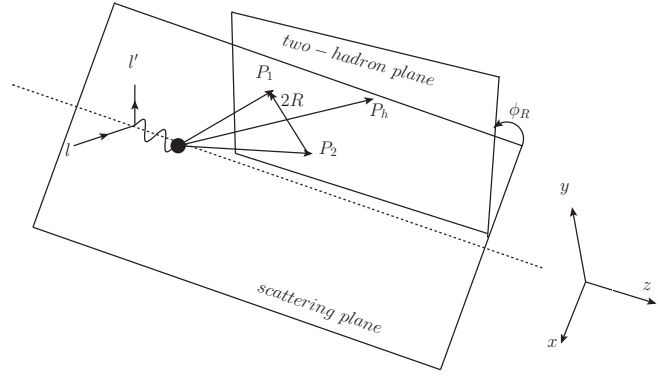}
  \caption{Angle definitions involved in the measurement of the single longitudinal-spin asymmetry in deep-inelastic production of two hadrons in the current region.}
  \label{fig:1}
\end{figure}

As displayed in Fig.\ref{fig:1}, the process under study is the dihadron production in SIDIS off a longitudinally polarized proton target:
\begin{align}
\mu(\ell)+p^\rightarrow (P)\longrightarrow \mu(\ell^\prime)+h^+(P_1)+h^-(P_2)+X,
\end{align}
where the four-momenta of the incoming and the outgoing leptons are denoted by $\ell$ and $\ell'$, $P$ is the momentum of the target with mass $M$.
In this process, the active quark with momentum $p$ is struck by the virtual photon with momentum $q=\ell- \ell'$. The final-state quark with momentum $k=p+q$ then fragments into two final-state hadrons, $h^+$ and $h^-$, plus unobserved state $X$.
The momenta of the pair are denoted by $P_1$, $P_2$, respectively.
In order to express the differential cross section as well as to calculate the DiFFs, we adopt the following kinematical variables
\begin{align}
\label{eq:invariants}
&x=\frac{k^+}{P^+},\qquad y=\frac{P\cdot q}{P\cdot l},\qquad z_{i}=\frac{P_{i}^-}{k^-},\\
&z=\frac{P_{h}^-}{k^-}=z_1+z_2,\qquad Q^2=-q^2,\qquad s=(P+l)^2, \\
&P_h=P_1+P_2, \qquad R=(P_1-P_2)/2, \qquad M_h=\sqrt{P_h^2}.
\end{align}
Here, we have used the light-cone coordinates $a^\mu=(a^+, a^-, a_T)$, where $a^{\pm}=(a^0\pm a^3)/\sqrt{2}$ and $a_T$ is the transverse component of the vector.
Therefore, $x$ represents the longitudinal momentum fraction of the initial quark, $z_i$ is the longitudinal momentum fraction of hadron $h_i$ found in the fragmented quark.
Furthermore, $M_h$, $P_h$ and $R$ are the invariant mass, the total momentum and the relative momentum of the hadron pair, respectively.

The momenta $P_h^{\mu}$, $k^{\mu}$ and $R^{\mu}$ thus can be written as~\cite{Bacchetta:2006un}
\begin{eqnarray}
P_h^{\mu}&=&\left[P^-_h,\frac{M_h^2}{2P^-_h},\vec{0} \right],\cr
k^{\mu}&=&\left[\frac{P_h^-}{z},\frac{z(k^2+\vec{k}_T^2)}{2P_h^-},\vec{k}_T \right],\cr
R^{\mu}&=&\left[\frac{|\vec{R}|P^-_h}{M_h}\cos\theta,-\frac{|\vec{R}|M_h}{2P^-_h}\cos\theta,|\vec{R}|\sin\theta\cos\phi_R,|\vec{R}|\sin\theta\sin\phi_R \right]\cr
&=&\left[\frac{|\vec{R}|P^-_h}{M_h}\cos\theta,-\frac{|\vec{R}|M_h}{2P^-_h}\cos\theta,\vec{R}_T^x,\vec{R}_T^y \right],
\end{eqnarray}
where $\theta$ is defined as the polar angle between the direction of $P_1$ in the center of mass frame of the hadron pair and the direction of $P_h$ in the lab frame~\cite{hep-ph/0212300},
$|\vec{R}|=\sqrt{M^2_h/4-m_\pi^2}$ and $\phi_R$ is the angle between the lepton plan and the two-hadron plane, as shown in Fig.\ref{fig:1}.
There are several useful expression of the scalar products as follows
\begin{align}
&P_h\cdot R= 0,\\
&P_h\cdot k= \frac{M_h^2}{2z}+z\frac{k^2+|\vec{k}_T|^2}{2},\\
&R\cdot k= (\frac{M_h}{2z}-z\frac{k^2+|\vec{k}_T|^2}{2M_h})|\vec{R}|\cos\theta-\vec{k}_T\cdot\vec{R}_T .
\end{align}

In the following we will consider the condition the lepton beam is unpolarized and the nucleon target is longitudinally polarized.
In this case, if the transverse momentum of the dihadron is integrated out, the differential cross section for an unpolarized (spin-averaged) target and a longitudinally polarized target can be cast to
\begin{align}
\frac{d^6\! \sigma^{}_{UU}}{d\cos \theta\;dM_h^2\;d\phi_R\;dz\;dx\;dy} &=  \frac{\alpha^2}{ Q^2 y}\,\left(1-y+\frac{y^2}{2}\right)   \sum_q e_a^2 f_1^a(x)\, D_1^a\bigl(z, M_{hh}^2, \cos \theta\bigr) , \label{eq:crossUU} \\
\frac{d^6\! \sigma^{}_{UL}}{d\cos \theta\;dM_h^2\;d\phi_R\;dz\;dx\;dy} &= -\frac{\alpha^2}{Q^2 y}\, S_L \,2(1-y)\sqrt{2-y} \sum_a e_a^2  {M\over Q}\frac{|\bm R| }{M_h} \, \sin \theta \, \sin\phi_{R}^{}  \nonumber\\
 &  \times
   \left[xh^a_L(x)H_{1}^{\sphericalangle,a}\bigl(z,M_h^2,\cos\theta\bigr)+\frac{M_h}{Mz} g_1(x)\widetilde{G}^{\sphericalangle,a}\bigl(z,M_h^2,\cos\theta\bigr)\right].
\label{eq:crossUL}
\end{align}
Here, the first subscript and the second subscript in $\sigma_{XY}$ denote the polarization states of the beam and the target, respectively.
In Eq.~(\ref{eq:crossUU}), $f_1^a(x)$ and $D_1^a\bigl(z, M_{hh}^2, \cos \theta\bigr)$ are the unpolarized PDF and unpolarized DiFF for flavor $a$; while in Eq.~(\ref{eq:crossUL}), $h_L^a(x)$ is the twist-3 distribution coupled with the chiral-odd DiFF $H_{1}^{\sphericalangle,a}\bigl(z,M_h^2,\cos\theta\bigr)$, $g_1^a(x)$ is the helicity distribution coupled with the twist-3 T-odd DiFF $\widetilde{G}^{\sphericalangle,a}\bigl(z,M_h^2,\cos\theta\bigr)$.
As shown in Eq.~(\ref{eq:crossUL}), both of these couplings contribute to the $\sin\phi_R$ azimuthal asymmetry in SIDIS.

The DiFFs $D_1^a$ and $H_{1}^{\sphericalangle,a}$ are encoded in the integrated quark-quark correlator $\Delta (z, R)$ for fragmentation~\cite{Bacchetta:2003vn}:
\begin{align}
  \Delta (z, R) &= z^2 \sumint \;
  \int \frac{d \xi^+}{2\pi} \; e^{\ii k\cdot\xi}\;
  \langle 0|\, U^+_{[0,\xi]} \, \psi(\xi) \,|P_h,R; X\rangle
  \langle X; P_h,R|\, \overline{\psi}(0)\,|0\rangle
  \Big|_{\xi^- = \vec \xi^{}_T  =0} \;  \nonumber\\
  &={1\over 16\pi} \left\{D_1 n\!\!\!/_-+ H_{1}^{\sphericalangle}{i\over M_h[R\!\!\!/_T,n\!\!\!/_-]}\right\},
\end{align}
where $U^a_{[b,c]}$ is the gauge-link running from $b$ to $c$ along $a$ to ensure the gauge invariance of the operator.

The twist-3 DiFF $\widetilde{G}^{\sphericalangle}$ arises from the multiparton correlation during the quark fragmentation, described by the quark-gluon-quark correlator~\cite{Bacchetta:2003vn,Lu:2015wja}:
\begin{eqnarray}
\widetilde{\Delta}_A^{\alpha}(z,k_T,R)&=&\frac{1}{2z}\sum_X\int\frac{d\xi^+d^2\xi_T}{(2\pi)^3}
e^{ik\cdot\xi}\langle0|\int_{\pm\infty^+}^{\xi^+}d\eta^+
{\cal{U}}^{\xi_T}_{(\infty^+,\xi^+)}\cr
&&\times g F_\bot^{-\alpha}{\cal{U}}^{\xi_T}_{(\eta^+,\xi^+)}\psi(\xi)|P_h,R;X\rangle\langle P_h,R;X|\bar{\psi}(0){\cal{U}}^{0_T}_{(0^+,\infty^+)}{\cal{U}}^{\infty^+}_{(0_T,\xi_T)}
|0\rangle\mid_{\eta^+=\xi^+=0,\eta_T=\xi_T}. \label{eq:deltaG}
\end{eqnarray}
Here, $F_\bot^{-\alpha}$ is the field strength tensor of the gluon.
After integrating out $\vec{k}_T$, one obtains
\begin{align}
\widetilde{\Delta}_A^{\alpha}(z,\cos \theta,M_h^2,\phi_R)=\frac{z^2|\vec{R}|}{8M_h}\int d^2\vec{k}_T \widetilde{\Delta}_A^{\alpha}(z,k_T,R).
\end{align}

The DiFF $\widetilde{G}^{\sphericalangle}$ thus can be extracted from $\widetilde{\Delta}_A^{\alpha}(z,k_T,R)$ by the trace
\begin{align}
\frac{\epsilon_T^{\alpha\beta}R_{T\beta}}{z}\widetilde{G}^{\sphericalangle}(z,\cos \theta,M_h^2)=4\pi\textrm{Tr}[\widetilde{\Delta}_A^{\alpha}(z,\cos \theta,M_h^2,\phi_R)\gamma^{-}\gamma_5].
\end{align}

As shown in Ref.~\cite{Bacchetta:2006un}, The DiFFs $D_1$ and $H_{1}^{\sphericalangle}$ can be expanded in the relative partial waves of the hadron pair system:
\begin{align}
&D^a_{1}(z,\cos \theta,M_h^2)=D^a_{1,oo}(z,M_h^2)+D^a_{1,ol}(z,M_h^2)\cos \theta+D^a_{1,ll}(z,M_h^2)(3\cos^2\theta-1),\label{eq:partial1}\\
&H_{1}^{\sphericalangle a}(z,\cos \theta, M_h^2)=H_{1,ot}^{\sphericalangle a}(z,M_h^2)+H_{1,lt}^{\sphericalangle a}(z,M_h^2)\cos \theta,
\label{eq:partial2}
\end{align}
in which the expansion is truncated at the $p$-wave level.
Similarly to $H_{1}^{\sphericalangle}$, we can also expand the twist-3 DiFF $\widetilde{G}^{\sphericalangle}$ up to the $p$-wave level as
\begin{align}
\widetilde{G}^{\sphericalangle}(z,\cos \theta, M_h^2)=\widetilde{G}^{\sphericalangle}_{ot}(z,M_h^2)+\widetilde{G}^{\sphericalangle}_{lt}(z,M_h^2)\cos \theta.
\end{align}
Here, $\widetilde{G}^{\sphericalangle}_{ot}$ originates from the interference of $s$ and $p$ waves, and $\widetilde{G}^{\sphericalangle}_{lt}$ comes from the interference of two $p$ waves with different polarization.

In this work, following the similar procedure in Ref.~\cite{Bacchetta:2002ux}, we will not consider the $\cos \theta$-dependent terms in the expansion of DiFFs because of two reasons. Firstly, $\cos \theta$-dependent terms correspond to the higher order contribution in the partial wave expansion and can only be significant when the two hadrons produce via a spin-1 resonance.
Secondly, when integrating out the angular $\theta$ in the interval $[-\pi, \pi]$ which is our case, the $\cos\theta$-dependent terms should vanish.
Therefore, we focus on the functions $D^a_{1,oo}$, $H_{1,ot}^{\sphericalangle}$ and  $\widetilde{G}^{\sphericalangle}_{ot}$.
In this scenario,
the ${\sin\phi_R}$ asymmetry of dihadron production in the single longitudinally polarized SIDIS may be expressed as~\cite{S.Sirtl},
\begin{align}
A^{\sin\phi_R}_{UL}(x,z,M_h^2) = -\frac{\sum_{a}e^2_a\frac{|\vec{R}|}{Q}
\left[\frac{|M|}{M_h}xh^a_L(x)H_{1,ot}^{\sphericalangle,a}(z,M_h^2)+\frac{1}{z} g_1(x)\widetilde{G}^{\sphericalangle}_{ot}(z,M_h^2)\right]}{\sum_{a}e^2_a f^a_1(x) D^a_{1,oo}(z,M_h^2)}.
\label{eq:AsinphiR}
\end{align}
Following the COMPASS convention, the depolarization factors are not included in the numerator and denominator.

\section{Model calculation of $\widetilde{G}^{\sphericalangle}_{ot}$}

\label{Sec.Model calculation}

Before actually calculating the unknown DiFF $\widetilde{G}^{\sphericalangle}_{ot}$ in the spectator model, we briefly review the calculation of twist-2 DiFFs $D_{1,oo}$ and $H_{1,ot}^{\sphericalangle}$ in the same model in Ref.~\cite{Bacchetta:2006un}.
In that paper, the twist-2 DiFFs $D_{1}$ and $H_{1}^{\sphericalangle}$ were expanded in terms of the relative partial waves of the hadron pair system up to
the $p$-wave level, similar to the expressions in Eqs.~(\ref{eq:partial1}), (\ref{eq:partial2}).
The DiFF $D_{1,oo}$ receives the pure $s$-wave contribution~(terms with vertex $|F^s|^2$) and the pure $p$-wave contribution~(terms with vertex $|F^p|^2$),
For $H_{1,ot}^{\sphericalangle}$ the gauge link (gluon exchange) would not contribute and therefore one
needs vertex factors ($F^s$ and $F^p$) that are complex.
Thus the DiFF $H_{1,ot}^{\sphericalangle}$ was calculated from the the interference of $s$ and $p$ waves which is proportional to $\textrm{Im}(F^{s\ast} F^p)$ or $\textrm{Im}(F^{s} F^{p\ast})$~\cite{Bacchetta:2006un}.
Here, the interference of $F^s$ and $F^p$ generates the necessary phase for the nonzero $H_{1,ot}^{\sphericalangle}$.
Furthermore, in the model the background of $s$ wave was assumed to be free of the resonances which means the vertex $F^s$ is real while the $p$-wave amplitude contains the contributions from the $\rho$ and the $\omega$ mesons.
By fitting the output of the PYTHIA Monte Carlo generator for the dihadron count in SIDIS, the parameters of the spectator model were fixed and the numerical results of the twist-2 DiFFs $D_{1,oo}$ and $H_{1,ot}^{\sphericalangle}$ were given.

\begin{figure}
  \centering
  \includegraphics[width=0.4\columnwidth]{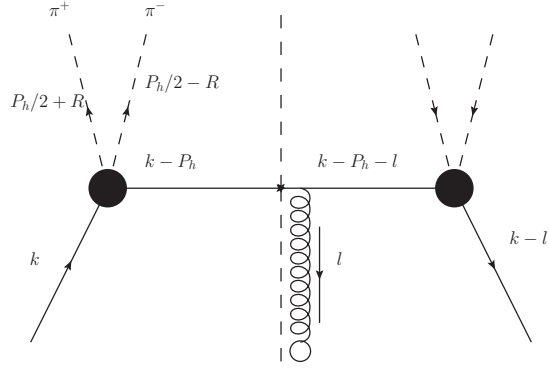}
  \caption{Diagrammatic representation of the correlation function $\widetilde{\Delta}_A^{\alpha}$ in the spectator model.}
  \label{fig:2}
\end{figure}

In the following, we present the calculation of the DiFF $\widetilde{G}^{\sphericalangle}_{ot}$ in the same spectator model.
The DiFF $\widetilde{G}^{\sphericalangle}_{ot}$ comes from the quark-gluon-quark correlation at the twist-3 level.
Here, different from the calculation of the twist-2 T-odd distribution $H_{1,ot}^{\sphericalangle}$ in Ref.~\cite{Bacchetta:2006un}, the gluon degree of freedom shows up explicitly, as given in the operator definition (\ref{eq:deltaG}). The corresponding diagram for the calculation in the spectator model is shown in Fig.~\ref{fig:2}. The left hand side of Fig.~\ref{fig:2} corresponds to the quark-hadron vertex $\langle P_{h};X|\bar{\psi}(0)|0\rangle $, while the right hand side corresponds to the vertex containing gluon rescattering $\langle 0|igF_{\perp}^{-\alpha}(\eta^{+})\psi(\xi^{+})|P_{h};X\rangle$.
Here, we apply the Feynman gauge, in which the transverse gauge links ${\cal U}^{ \bm \xi_T}$ and ${\cal U}^{\bm 0_T}$ can be neglected~\cite{Ji:2002aa,Belitsky:2002sm}.

Therefore, the $s$ and $p$ wave contributions to the quark-gluon-quark correlator for dihardon fragmentation in the spectator model can be written as
\begin{eqnarray}
&&\widetilde{\Delta}_A^{\alpha}(k,P_h,R)=i\frac{C_F\alpha_s}{2(2\pi)^2(1-z)P_h^-}\frac{1}{k^2-m^2}\int \frac{d^4l}{(2\pi)^4}(l^-g_T^{\alpha\mu}-l^\alpha_T g^{-\mu})\cr
&&\frac{(k\!\!\!/ -l\!\!\!/ +m)(F^{s\star} e^{-\frac{k^2}{\Lambda_s^2}}+F^{p\star} e^{-\frac{k^2}{\Lambda_p^2}} R\!\!\!/)(k\!\!\!/ -P\!\!\!/_h-l\!\!\!/+m_s)\gamma_{\mu}(k\!\!\!/ -P\!\!\!/_h+m_s)(F^{s} e^{-\frac{k^2}{\Lambda_s^2}}+F^{p} e^{-\frac{k^2}{\Lambda_p^2}} R\!\!\!/)(k\!\!\!/+m)}{(-l^-\pm i\epsilon)((k-l)^2-m^2-i\epsilon)((k-P_h-l)^2-m_s^2-i\epsilon)(l^2-i\epsilon)}, \label{Delta1}
\end{eqnarray}
where $m$ and $m_s$ are the masses of the fragmenting quark and the spectator, and where the factor $(l^-g_T^{\alpha \mu} -l_T^\alpha g^{-\mu})$ comes from the Feynman rule corresponding to the gluon field strength tensor, as denoted by the open circle in Fig.~\ref{fig:2}. $F^s$, $F^p$ are the vertices refer to the $s$-wave contribution and $p$-wave contribution~\cite{Bacchetta:2006un} and have the following forms:
\begin{align}
F^s& = f_s \,,\\
F^p& = f_\rho \frac{(M_h^2-M_\rho^2)-i\Gamma_\rho M_\rho}{(M_h^2-M_\rho^2)+\Gamma_\rho^2 M_\rho^2}+f_\omega \frac{(M_h^2-M_\omega^2)-i\Gamma_\omega M_\omega}{(M_h^2-M_\rho^2)+\Gamma_\omega^2 M_\omega^2}\nonumber\\
&-if'_\omega \frac{\sqrt{\hat{\lambda}(M_\omega^2,M_h^2,m_\pi^2)}\Theta(M_\omega-m_\pi-M_h)}
{4\pi\Gamma_\omega^2[4M_\omega^2m_\pi^2+\hat{\lambda}(M_\omega^2,M_h^2,m_\pi^2)]^{\frac{1}{4}}}\,.
\end{align}
Here, $\hat{\lambda}(M_\omega^2,M_h^2,m_\pi^2)=(M_\omega^2-(M_h+m_\pi)^2)(M_\omega^2-(M_h-m_\pi)^2)$, and $\Theta$ denotes the unit step function.
The first two terms of $F^p$ can be identified with the contributions of the $\rho$ and the $\omega$ resonances decaying into two pions.
The masses and widths of the two resonances are adopted from the PDG~\cite{Eidelman}: $M_\rho = 0.776 \mathrm{GeV}$, $\Gamma_\rho = 0.150 \mathrm{GeV}$, $M_\omega = 0.783 \mathrm{GeV}$, $\Gamma_\omega = 0.008 \mathrm{GeV}$.

In Eq.~(\ref{Delta1}), in principle one of the exponential from factors should depend on the loop momentum $l$. Here we follow the choice in Ref.~\cite{Bacchetta:2007wc} to replace $(k-l)^2$ in the form factor to $k^2$.
The reason for this choice is that the form factor is introduced to the purpose of cutting
off the high-$k_T$ region. Maintain the form factor depending only on $k^2$ can pull the form factor out of the integral and simplify the calculation.

Expanding Eq.~(\ref{Delta1}), we arrive at
\begin{eqnarray}
\label{Delta}
&&\widetilde{\Delta}_A^{\alpha}(z,\cos \theta,M_h^2,\phi_R)=i\frac{C_F\alpha_s z^2|\vec{R}|}{16(2\pi)^5(1-z)M_h P_h^-}\int d|\vec{k}_T|^2\int d^4l\frac{l^-g_T^{\alpha\mu}-l^\alpha_T g^{-\mu}}{k^2-m^2}\cr
&&\bigg{[}|F^{s}|^2 e^{-\frac{2k^2}{\Lambda_s^2}}\frac{(k\!\!\!/ -l\!\!\!/ +m)(k\!\!\!/ -P\!\!\!/_h-l\!\!\!/+m_s)\gamma_{\mu}(k\!\!\!/ -P\!\!\!/_h+m_s)(k\!\!\!/+m)}{(-l^-\pm i\epsilon)((k-l)^2-m^2-i\epsilon)((k-P_h-l)^2-m_s^2-i\epsilon)(l^2-i\epsilon)}\cr
&&+|F^{p}|^2 e^{-\frac{2k^2}{\Lambda_p^2}}\frac{(k\!\!\!/ -l\!\!\!/ +m) R\!\!\!/(k\!\!\!/ -P\!\!\!/_h-l\!\!\!/+m_s)\gamma_{\mu}(k\!\!\!/ -P\!\!\!/_h+m_s)R\!\!\!/(k\!\!\!/+m)}{(-l^-\pm i\epsilon)((k-l)^2-m^2-i\epsilon)((k-P_h-l)^2-m_s^2-i\epsilon)(l^2-i\epsilon)}\cr
&&+(F^{s\star}F^{p}) e^{-\frac{2k^2}{\Lambda_{sp}^2}}\frac{(k\!\!\!/ -l\!\!\!/ +m)(k\!\!\!/ -P\!\!\!/_h-l\!\!\!/+m_s)\gamma_{\mu}(k\!\!\!/ -P\!\!\!/_h+m_s) R\!\!\!/(k\!\!\!/+m)}{(-l^-\pm i\epsilon)((k-l)^2-m^2-i\epsilon)((k-P_h-l)^2-m_s^2-i\epsilon)(l^2-i\epsilon)}\cr
&&+(F^{s}F^{p\star}) e^{-\frac{2k^2}{\Lambda_{sp}^2}}\frac{(k\!\!\!/ -l\!\!\!/ +m)R\!\!\!/(k\!\!\!/ -P\!\!\!/_h-l\!\!\!/+m_s)\gamma_{\mu}(k\!\!\!/ -P\!\!\!/_h+m_s) (k\!\!\!/+m)}{(-l^-\pm i\epsilon)((k-l)^2-m^2-i\epsilon)((k-P_h-l)^2-m_s^2-i\epsilon)(l^2-i\epsilon)}\bigg{]},
\end{eqnarray}
where $\Lambda_{s}$ and $\Lambda_{p}$ are the $z$-dependent $\Lambda$-cutoffs having the form~\cite{Bacchetta:2006un}
\begin{align}
\Lambda_{s,p} = \alpha_{s,p} z^{\beta_{s,p}} (1-z)^{\gamma_{s,p}},
\end{align}
and $2/\Lambda^2_{sp}=1/\Lambda^2_{s}+1/\Lambda^2_{p}$.
The on-shell condition of the spectator gives the relation between $k^2$ and the transverse momentum $\vec k_T$:~\cite{hep-ph/9907488},
\begin{eqnarray}
k^2&=&\frac{z}{1-z}|\vec{k}_T|^2+\frac{M_s^2}{1-z}+\frac{M_h^2}{z}.
\end{eqnarray}

The first and second lines of Eq.~(\ref{Delta}) provide the pure $s$-wave and $p$-wave contributions, respectively. Therefore, they will not contribute to the interference of $s$ and $p$-waves functions $\widetilde{G}^{\sphericalangle}_{ot}$, only the third and fourth lines have contribution.
At one loop level, in principle there are two sources for nonzero $\widetilde{G}^{\sphericalangle}_{ot}$. One is the imaginary part of the loop integral over $l$, combined with the real part of $F^{s*}F^p$.
The other is the imaginary part of $F^{s*}F^p$, combined with the real part of the loop integral over $l$.
For the imaginal part of the integral, we apply the Cutkosky cutting rules; while for the real part of the integral, we adopt the Feynman parametrization.
Thus, the final result for $\widetilde{G}^{\sphericalangle}_{ot}(z,M_h^2)$ has the form
\begin{align}
\widetilde{G}^{\sphericalangle}_{ot}(z,M_h^2)& =\frac{\alpha_s C_F z^2 |\vec{R}|}{8 (2\pi)^4 (1-z)M_h}\frac{1}{k^2-m^2}\int d|\vec{k}_T|^2 e^{-\frac{2 k^2}{\Lambda_{sp}^2}}
\bigg{\{}  \textrm{Im}(F^{s*}F^p) \,C  \nonumber\\
&+ \textrm{Re}(F^{s*}F^p)(k^2-m^2)m_s \big{[}(A + z B) - I_2 \big{]}  \bigg{\}}\,. \label{Gtilde}
\end{align}
Here,  the coefficients $A$ and $B$ come from the decomposition of the integral~\cite{Lu:2015wja,1607.01638},
\begin{align}
&\int d^4l { l^\mu\, \delta(l^2)\, \delta((k-l)^2-m^2)\over (k-P_h-l)^2-m_s^2}=A\, k^\mu + B\, P_h^\mu\,,
\end{align}
and have the expressions
\begin{align}
A&={I_{1}\over \lambda(M_h,m_s)} \left(2k^2 \left(k^2 - m_s^2 - M_h^2\right) {I_{2}\over \pi}+\left(k^2+M_h^2 - m_s^2\right)\right), \\
B&=-{2k^2 \over \lambda(M_h,m_s) } I_{1}\left (1+{k^2+m_s^2-M_h^2 \over \pi} I_{2}\right)\,.
\end{align}
The functions $I_{i}$ appearing in the above equations are defined as~\cite{Amrath:2005gv}
\begin{align}
I_{1} &=\int d^4l \delta(l^2) \delta((k-l)^2-m^2) ={\pi\over 2k^2}\left(k^2-m^2\right)\,, \\
I_{2} &= \int d^4l { \delta(l^2) \delta((k-l)^2-m^2)\over (k-P_h-l)^2-m_s^2}
={\pi\over 2\sqrt{\lambda(M_h,m_s)} }  \ln\left(1-{2\sqrt{ \lambda(M_h,m_s)}\over k^2-M_h^2+m_s^2 + \sqrt{ \lambda(M_h,m_s)}}\right)\,,
\end{align}
where $\lambda(M_h,m_s)=(k^2-(M_h+m_s)^2)(k^2-(M_h-m_s)^2)$.

The coefficient $C$ in Eq.~(\ref{Gtilde}) has the form
\begin{align}
&C=m\int_0^1 dx \int_0^{1-x} dy \frac{-2\left[(x+y)k\cdot P_h-y M_h^2\right] +(k^2-m^2) }{x(1-x)k^2+2k\cdot(k-P_h)xy+x m^2 +y^2 m_s^2}\,,
\end{align}
which is proportional to the fragmenting quark mass $m$.

\begin{figure*}
  \centering
  \includegraphics[width=0.48\columnwidth]{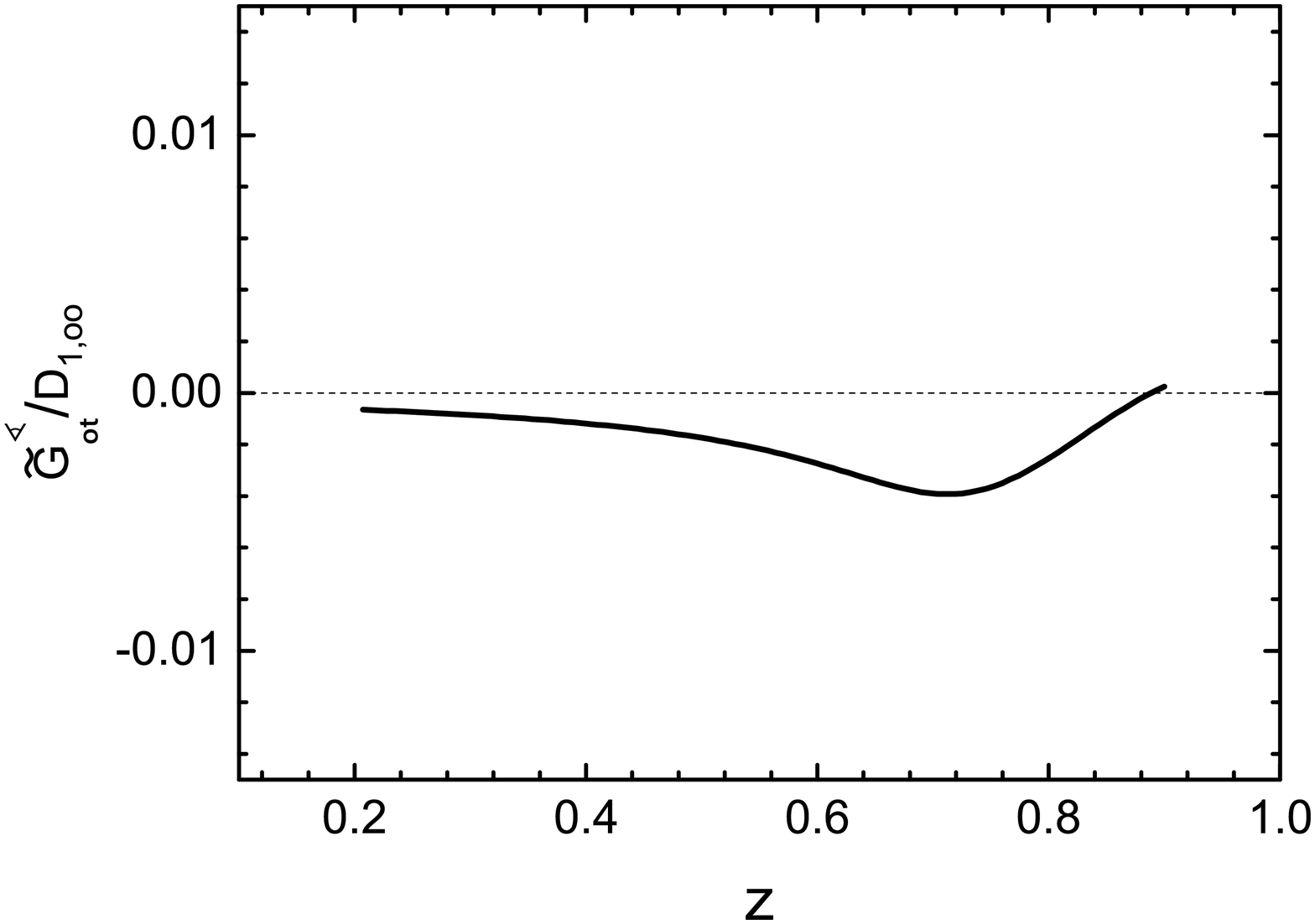}
  \includegraphics[width=0.48\columnwidth]{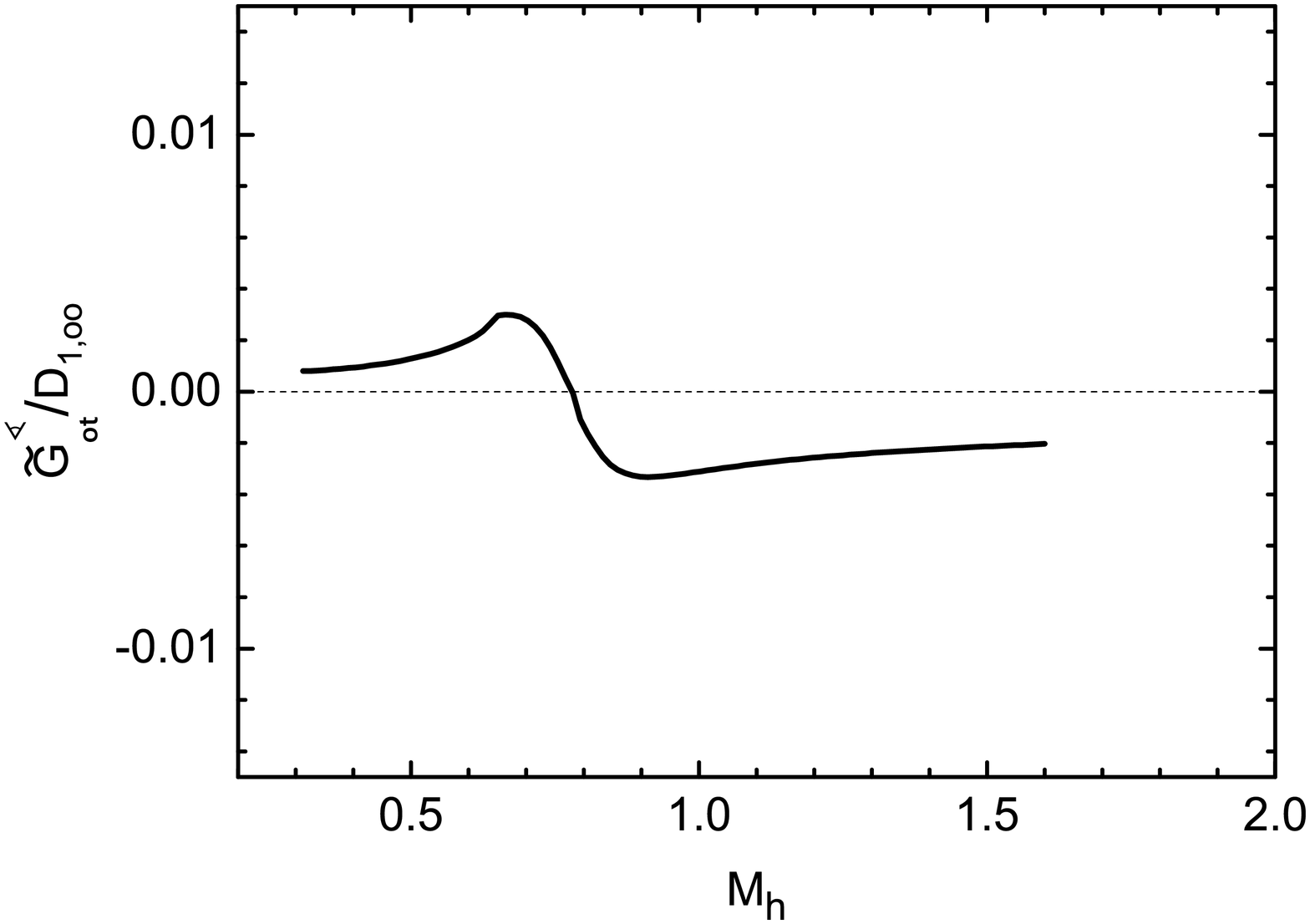}
  \caption{The twist-3 DiFF $\widetilde{G}^{\sphericalangle}_{ot}$ as the functions of $z$ (left panel) and $M_h$ (right panel) in the spectator model, normalized  by the unpolarized DiFF $D_{1,oo}$.}
  \label{fig:g}
\end{figure*}

\section{Numerical estimate}

\label{Sec.numerical}

In order to obtain the numerical result for $\widetilde{G}^{\sphericalangle}_{ot}(z,M_h^2)$, we choose the values for the parameters $m$, $m_s$, $\alpha_{s,p}$, $\beta_{s,p}$ and $\gamma_{s,p}$ from Ref.~\cite{Bacchetta:2006un}, where the model parameters were tuned to the output of PYTHIA event generator for dihadron production in SIDIS:
\begin{align}
\alpha_s& = 2.60 \mathrm{GeV}\,,\qquad  \beta_s = -0.751\,,\qquad \gamma_s = -0.193 \nonumber\,,\\
\alpha_p& = 7.07 \mathrm{GeV}\,,\qquad  \beta_p = -0.038\,,\qquad \gamma_p = -0.085 \nonumber\,,\\
f_s& = 1197 \mathrm{GeV}^{-1}\,,\qquad  f_\rho = 93.5\,,\qquad f_\omega = 0.63 \nonumber\,,\\
f'_\omega& = 75.2 \,,\qquad  M_s = 2.97 M_h\,,\qquad m = 0.0\mathrm{GeV} (\textrm{fixed})\,.
\end{align}
Particularly, in Ref.~\cite{Bacchetta:2006un} the quark mass $m$ is fixed as $0$ GeV.
Therefore, in this  model, only the $\textrm{Re}(F^{s*}F^p)$ term  in Eq.~(\ref{Gtilde}) contributes to $\widetilde{G}^{\sphericalangle}_{ot}$ numerically, since the $\textrm{Im}(F^{s*}F^p)$ term is proportional to the quark mass $m$. As for the strong coupling, we choose $\alpha_s\approx0.3$.

In the left panel of Fig.~\ref{fig:g}, we plot the DiFF $\widetilde{G}^{\sphericalangle}_{ot}$ (normalized by $D_{1,oo}$) as function of $z$ with $M_h$ integrated over the region $0.3 \textrm{GeV} <M_h <1.6 \textrm{GeV}$.
In the right panel of Fig.~\ref{fig:g}, we plot $\widetilde{G}^{\sphericalangle}_{ot}$ (normalized by $D_{1,oo}$) as function of $M_h$ with $z$ integrated over the region $0.2 <z <0.9$.
We find that $\widetilde{G}^{\sphericalangle}_{ot}$ is negative in the entire $z$ region when $M_h$ is integrated out, while it is positive in the region $M_h <0.8$ GeV and negative in the region $M_h >0.8$ GeV.
The size of $\widetilde{G}^{\sphericalangle}_{ot}$ is less than one percent compared to the leading-twist DiFF $D_{1,oo}$.

In the following, we numerically estimate the $\sin\phi_R$ azimuthal asymmetry in the dihadron production off a longitudinally polarized proton by considering both the $h_L H_{1,ot}^{\sphericalangle,a}$ term and the $g_1 \widetilde{G}^{\sphericalangle}_{ot}$ term.
Using Eq.~(\ref{eq:AsinphiR}), we can obtain the expressions of  the $x$-dependent, $z$-dependent and $M_h$-dependent $\sin\phi_R$ asymmetry as follows
\begin{align}
\label{Ax}
A^{\sin\phi_R}_{UL}(x)=-\frac{\int dz \int dM_h 2M_h \frac{|\vec{R}|}{Q}
[\frac{|M|}{M_h}(4 h_L^u(x)+h_L^d(x))xH_{1,ot}^{\sphericalangle}(z,M_h^2)+\frac{1}{z} (4 g_1^u(x)+g_1^d(x))\widetilde{G}^{\sphericalangle}_{ot}(z,M_h^2)]
}{\int dz \int dM_h 2M_h (4 f_1^u(x)+f_1^d(x))D_{1,oo}(z,M_h^2)},\\
\label{Az}
A^{\sin\phi_R}_{UL}(z)=-\frac{\int dx\int dM_h 2M_h \frac{|\vec{R}|}{Q}
[\frac{|M|}{M_h}(4 h_L^u(x)+h_L^d(x))xH_{1,ot}^{\sphericalangle}(z,M_h^2)+\frac{1}{z} (4 g_1^u(x)+g_1^d(x))\widetilde{G}^{\sphericalangle}_{ot}(z,M_h^2)]
}{\int dx \int dM_h 2M_h (4 f_1^u(x)+f_1^d(x))D_{1,oo}(z,M_h^2)},\\
\label{Amh}
A^{\sin\phi_R}_{UL}(M_h)=-\frac{\int dx\int dz \frac{|\vec{R}|}{Q}
[\frac{|M|}{M_h}(4 h_L^u(x)+h_L^d(x))xH_{1,ot}^{\sphericalangle}(z,M_h^2)+\frac{1}{z} (4 g_1^u(x)+g_1^d(x))\widetilde{G}^{\sphericalangle}_{ot}(z,M_h^2)]
}{\int dx\int dz (4 f_1^u(x)+f_1^d(x))D_{1,oo}(z,M_h^2)}.
\end{align}

\begin{figure}
  \centering
  \includegraphics[width=0.48\columnwidth]{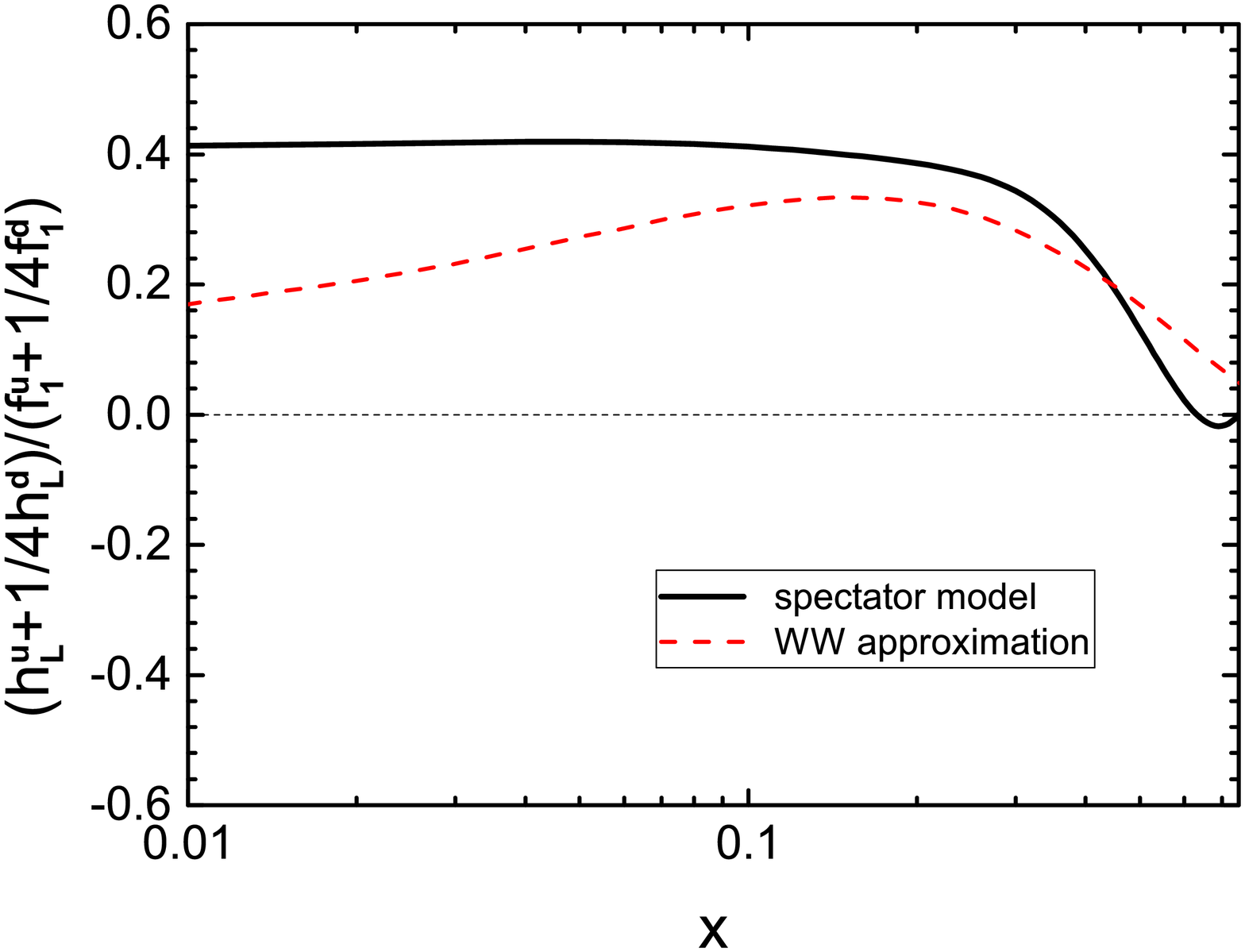}
  \caption{The twist-3 distribution function normalized by the unpolarized distribution $(h_L^u+1/4h_L^d)/(f_1^u+1/4f_1^d)$ as function of $x$. The solid curve corresponds to the spectator model result, while the dashed curve denotes the result from the Wandzura-Wilczek for $h_L$.}
  \label{fig:hLcom}
\end{figure}

\begin{figure*}
  \centering
  \includegraphics[width=0.48\columnwidth]{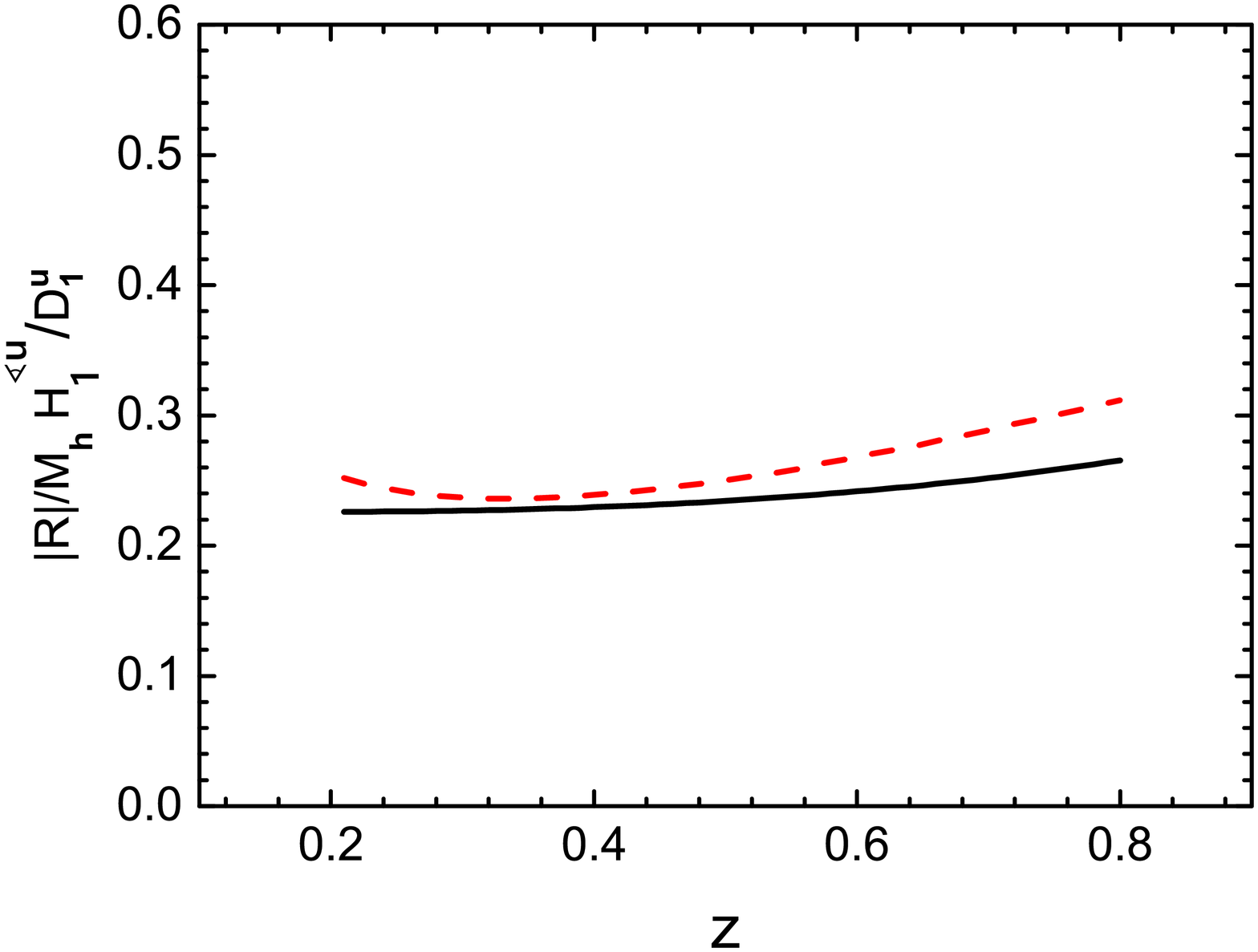}
  \includegraphics[width=0.48\columnwidth]{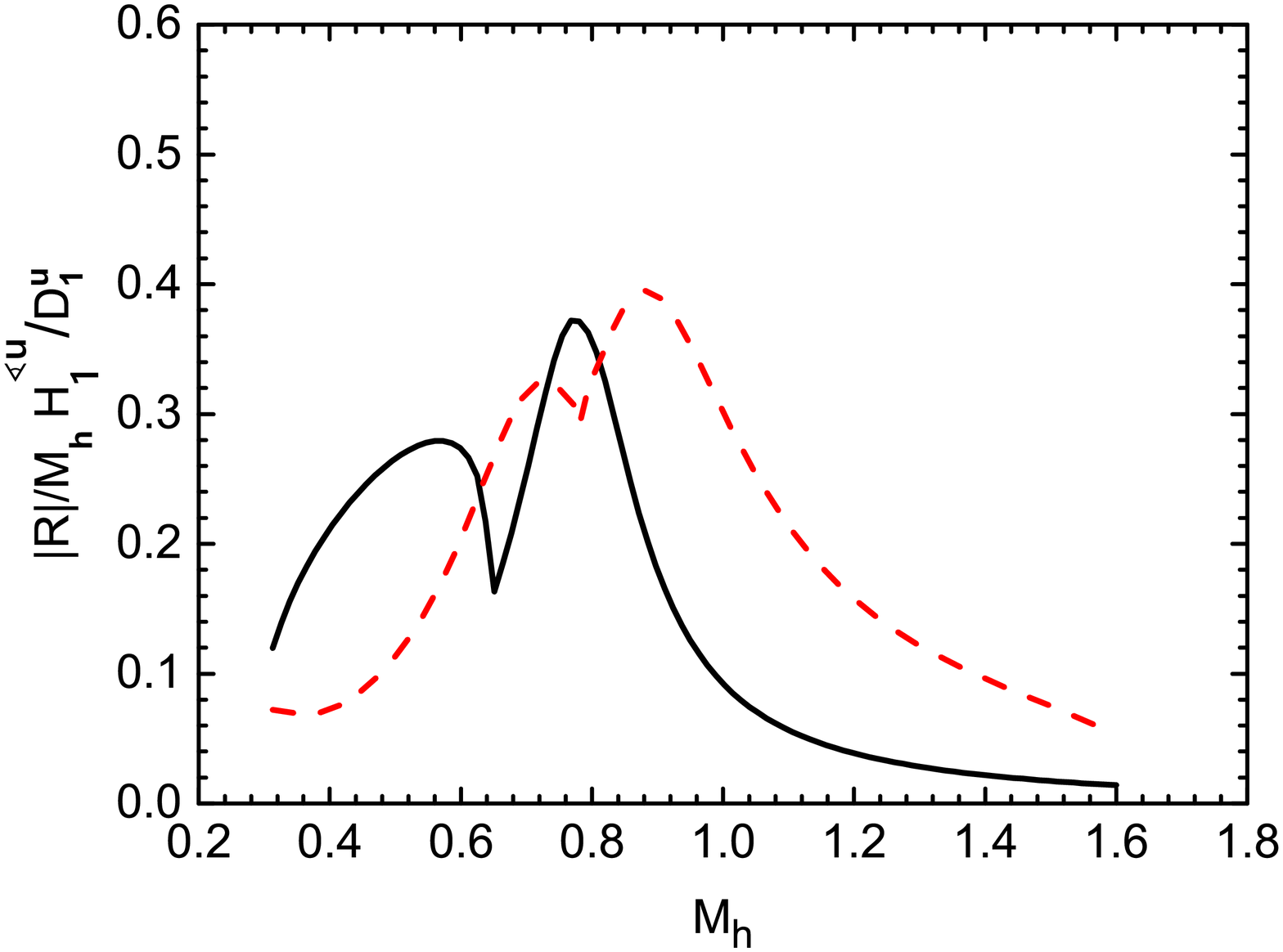}
  \caption{The ratio of the DiFFs $H_{1}^{\sphericalangle,u}$ and $D_{1}^u$ as the functions of $z$ (left panel) and $M_h$ (right panel). The dashed curves correspond to result from parametrizations for DiFFs, and the solid curve corresponds to spectator model result.}
  \label{fig:H1}
\end{figure*}

For the other DiFFs $H_{1,ot}^{\sphericalangle,a}(z,M_h^2)$ and $D_{1,oo}(z,M_h^2)$ needed in the calculation, we apply the same spectator model results from Ref.~\cite{Bacchetta:2006un}. For the twist-3 distribution $h_L$, we choose the result in Ref.~\cite{Lu:1404.4229}, as for the twist-2 PDFs $f_1$ and $g_1$, we adopt the results calculated from the same model~\cite{Bacchetta:2008af} for consistency.

As there are several model inputs in the expression of the $\sin\phi_R$ asymmetry, it is necessary to check the reliability of the model calculations of the distributions and DiFFs.
For the twist-3 distribution function $h_L(x)$, we compare the spectator model result in Ref.~\cite{Lu:1404.4229} with that from the Wandzura-Wilczek approximation~\cite{Jaffe:1992}
\begin{align}
h_L(x)\approx 2x\int_x^1\frac{dy}{y^2}h_1(y).
\label{eq:WW}
\end{align}
Here, $h_1(x)$ is the transversity distribution function and has been extracted~\cite{Anselmino:2008jk} phenomenologically.
It is known that for the twist-3 distribution $g_T(x)$ the Wandzura-Wilczek approximation works reasonably well~\cite{Wandzura:1977qf}.
In Fig.~\ref{fig:hLcom}, we plot the numerical results of $(h_L^u+h_L^d/4)/(f_1^u+f_1^d/4)$ which is the relevant one appearing in the expression for the $\sin\phi_R$ asymmetry. The solid line denotes the model result, while the dashed line corresponds to the result from the Wandzura-Wilczek approximation.
The comparison shows that the former one qualitatively agrees with the latter one.

Secondly, in order to check the model input for the DiFFs, we also compare the spectator model result of the DiFF $H_{1}^{\sphericalangle,a}$ with the parameterized result extracted in Ref.~\cite{Courtoy:2012ry}.
In Fig.~\ref{fig:H1}, we compare the ratio $|R|/M_h\,H_{1}^{\sphericalangle,u} / D_{1}^u$ from the model (solid lines) and from the parametrization (dashed lines) as functions of $z$ (left panel) and $M_h$ (right panel), respectively.
For the $z$-dependent result we have integrated over $M_h$ in the region $0.3 \textrm{GeV} <M_h <1.6 \textrm{GeV}$, while the $M_h$-dependent result we have integrated over $z$ in the region $0.2 <z <0.9$.
A better agreement is found in the $z$-dependent result.

To compare estimate the $\sin\phi_R$ asymmetry in SIDIS at COMPASS, we adopt the following kinematical cuts~\cite{S.Sirtl}
\begin{align}
&0.003<x<0.4,\quad 0.1<y<0.9, \quad 0.2<z<0.9,\nonumber\\
&0.3 \textrm{GeV} <M_h<1.6\textrm{GeV}, \quad Q^2> 1\mathrm{GeV}^2, \quad W>5\mathrm{GeV}.
\label{eq:cuts}
\end{align}
Here, $W$ is invariant mass of the virtual photon-nucleon system and $W^2=(P+q)^2\approx\frac{1-x}{x}Q^2$.
We note that we have not considered the evolution effect of the distribution functions and DiFFs, as the evolution of the twist-3 functions $h_L$ and $\widetilde{G}^{\sphericalangle}_{ot}$ remain unknown.

\begin{figure*}
  \centering
  \includegraphics[width=0.32\columnwidth]{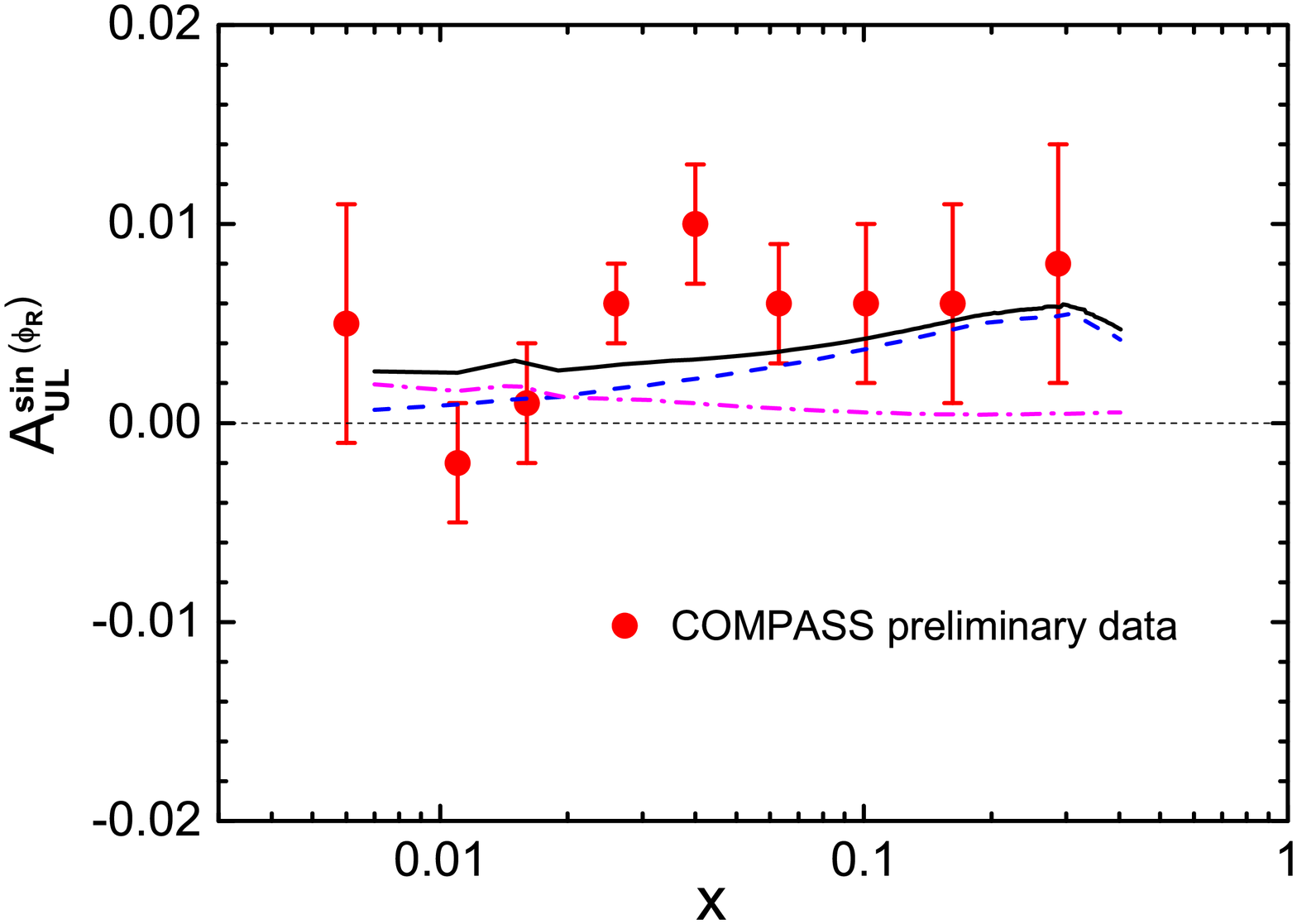}
  \includegraphics[width=0.32\columnwidth]{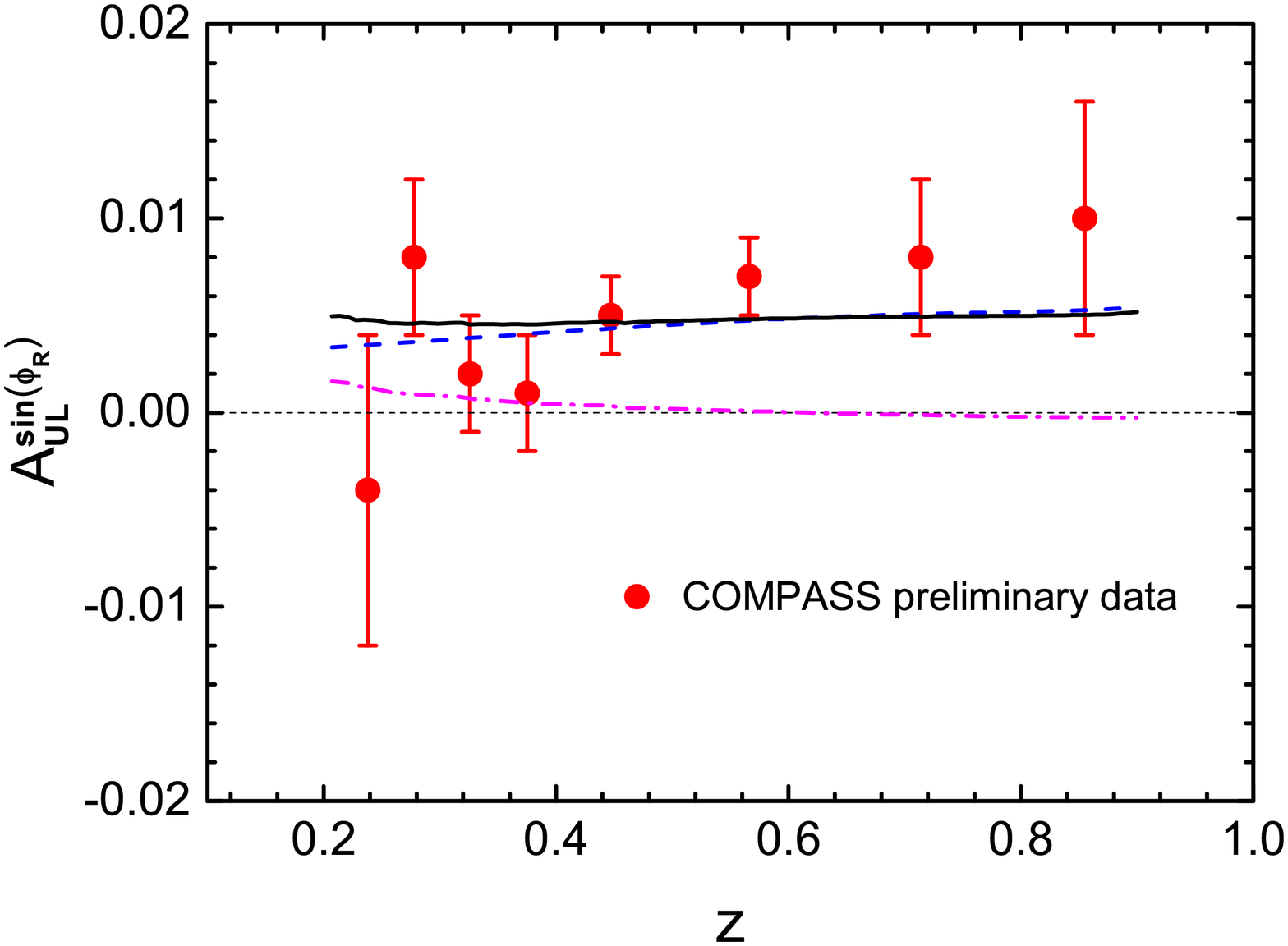}
  \includegraphics[width=0.32\columnwidth]{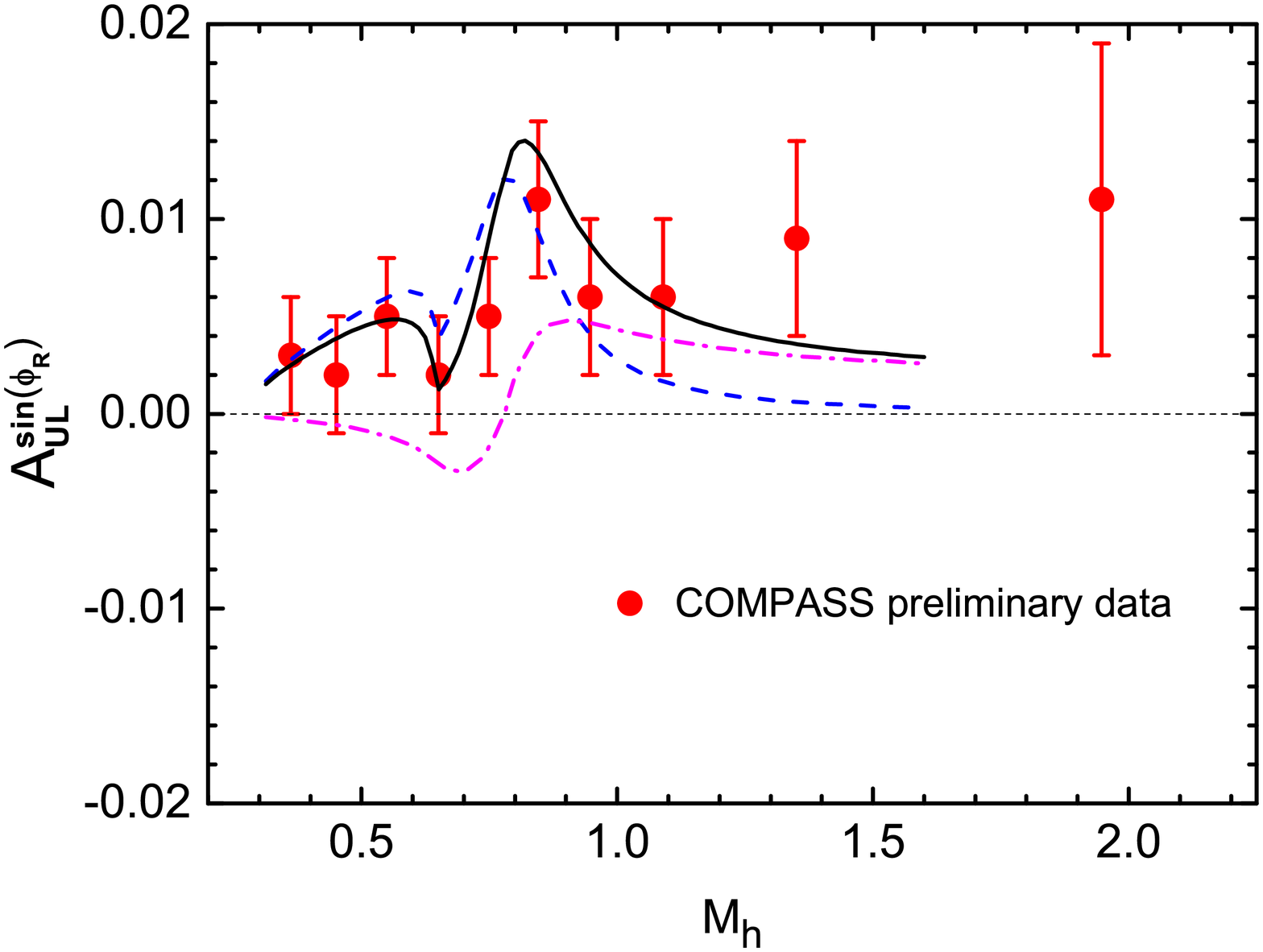}
  \caption{The $\sin\phi_R$ azimuthal asymmetry in dihadron production off the longitudinally polarized proton as functions of $x$ (left panel), $z$ (central panel) and $M_h$ (right panel) at COMPASS. The full circles show the COMPASS preliminary data~\cite{S.Sirtl} for comparison. The dashed curves denote the contribution from the $h_L\, H_{1,ot}^{\sphericalangle}$ term, the dashed-dotted curves represent the contribution from the $g_1\,\widetilde{G}^{\sphericalangle}$ term, and the solid lines display the sum of two contributions.}
  \label{fig:asy}
\end{figure*}

In Fig.~\ref{fig:asy}, we plot the $\sin\phi_R$ asymmetry in dihadron production off the longitudinally polarized proton at the kinematics of COMPASS.
The $x$-, $z$- and $M_h$-dependent asymmetries are depicted in the left panel, central, and right panels of the figure, respectively.
The dashed lines represent the contribution from the $h_L H_{1,ot}^{\sphericalangle,a}$ term and the dot-dashed lines depict the contribution from the $g_1\,\widetilde{G}^{\sphericalangle}$ term.
The solid lines display the sum of the two contributions.
The full circles with error bars show the preliminary measurement by the COMPASS collaboration for comparison.
We find that in the large $x$ region and in the small $M_h$ region, the contribution from the $h_L H_{1,ot}^{\sphericalangle,a}$ term dominates the asymmetry. The $g_1 \,\widetilde{G}^{\sphericalangle}$ becomes important in the small $x$ region and large $M_h$ region.
Combining the contributions from the two terms, our calculation agrees with the COMPASS preliminary data on the $\sin\phi_R$ asymmetry.
Particularly, in the region $M_h>1$ GeV, the contribution from the $h_L H_{1,ot}^{\sphericalangle,a}$ underestimates the asymmetry, whereas the inclusion of the $g_1 \,\widetilde{G}^{\sphericalangle}$ term can yield a better description of the COMPASS preliminary data.

\begin{figure*}
  \centering
  \includegraphics[width=0.32\columnwidth]{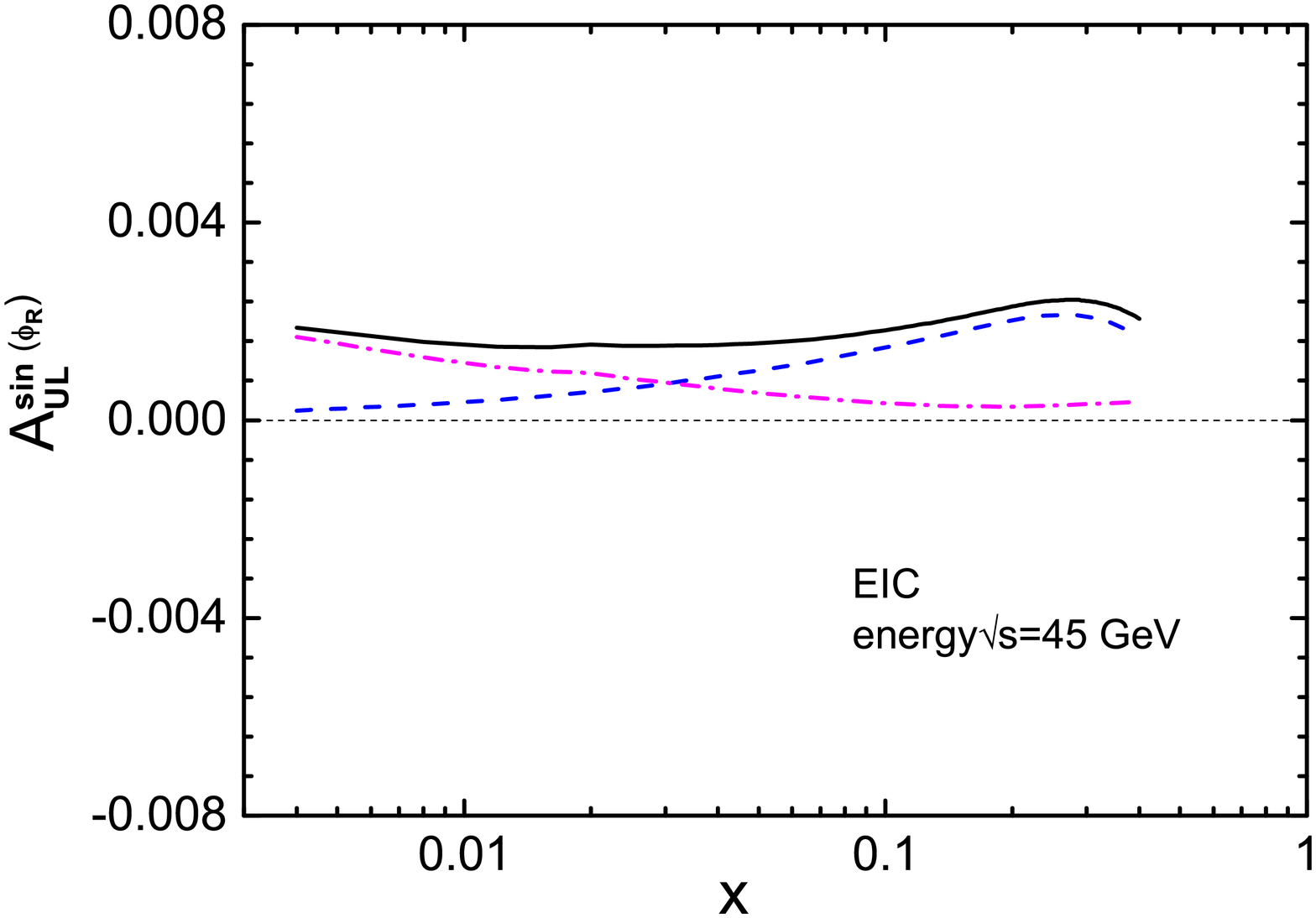}
  \includegraphics[width=0.32\columnwidth]{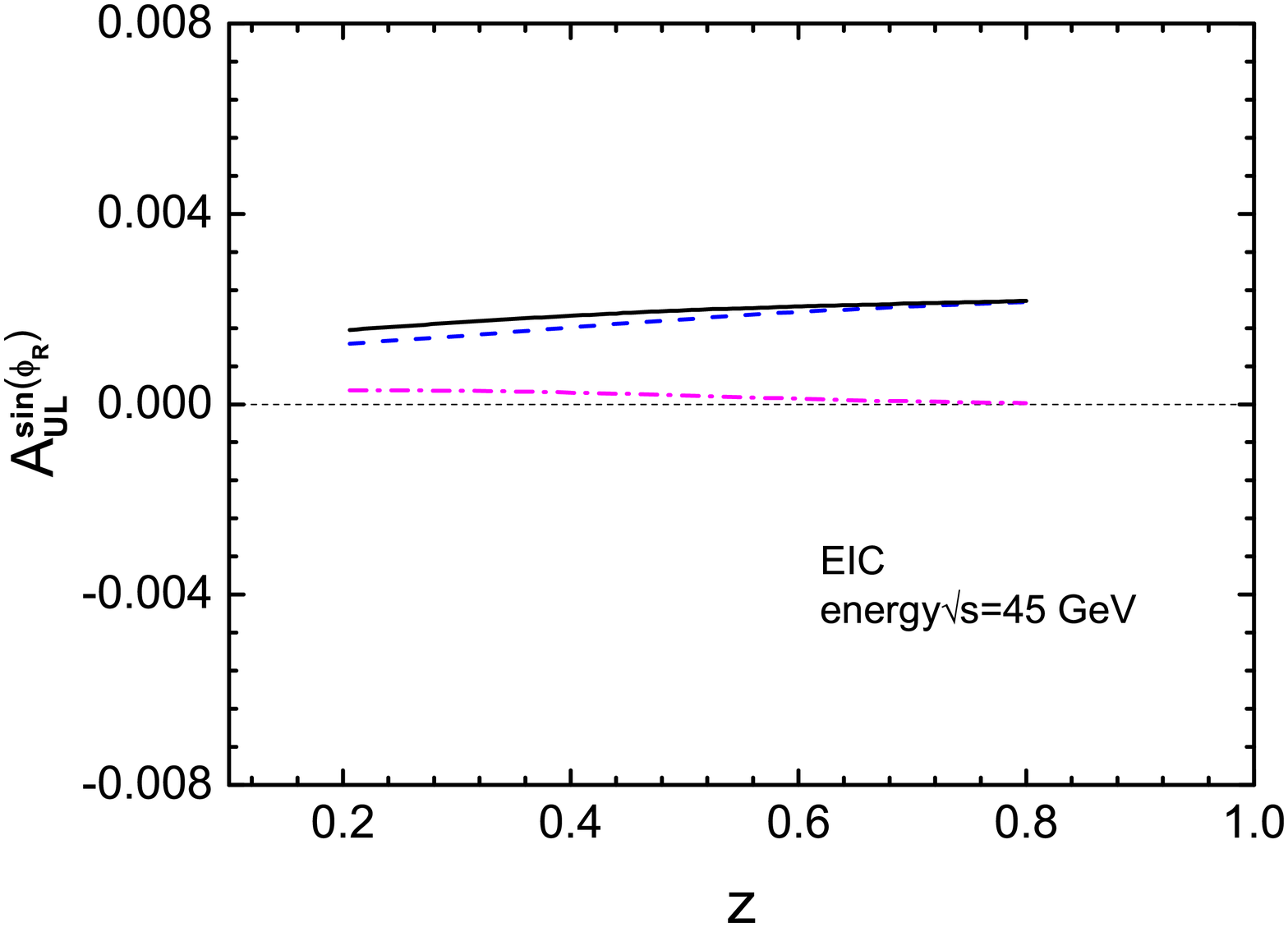}
  \includegraphics[width=0.32\columnwidth]{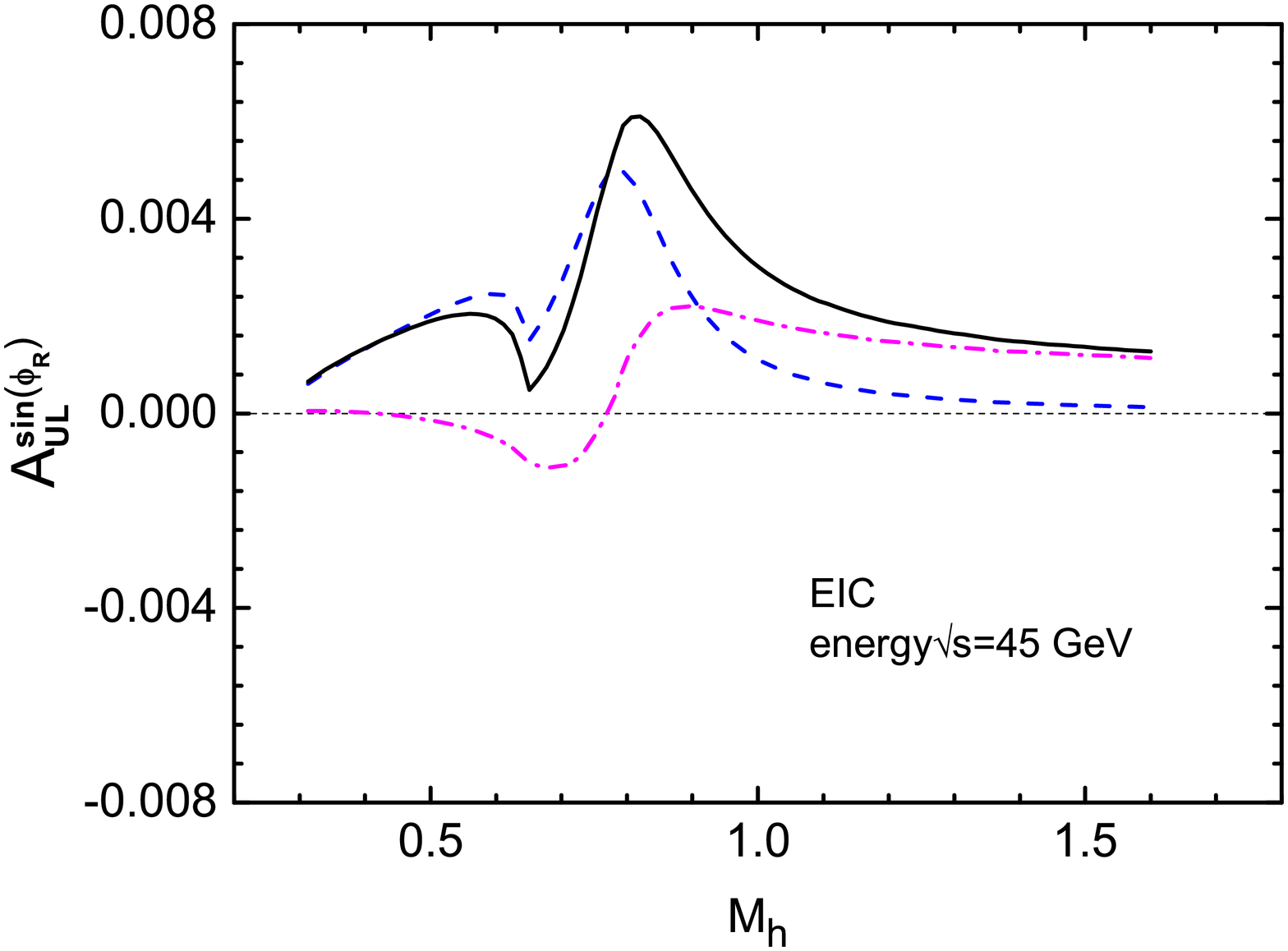}
  \caption{The $\sin\phi_R$ azimuthal asymmetry in dihadron production off the longitudinally polarized proton as functions of $x$ (left panel), $z$ (central panel) and $M_h$ (right panel) at the kinematics of the EIC. The dashed curves denote the contribution from the $h_L\, H_{1,ot}^{\sphericalangle}$ term, the dashed-dotted curves represent the contribution from the $g_1\,\widetilde{G}^{\sphericalangle}$ term, and the solid lines display the sum of two contributions.}
  \label{fig:asy1}
\end{figure*}

In addition, we also make the prediction on the $\sin\phi_R$ asymmetry in the single-longitudinally polarized SIDIS at the future EIC. Such a facility would be ideal to
study this observable.
We adopt the following EIC kinematical cuts~\cite{Matevosyan:2015gwa}
\begin{align}
&\sqrt{s}=45\mathrm{GeV}, \quad 0.001<x<0.4,\quad 0.01<y<0.95, \quad 0.2<z<0.8,\nonumber\\
&0.3 \textrm{GeV} <M_h<1.6\textrm{GeV}, Q^2>1\mathrm{GeV}^2,\quad \quad W>5\mathrm{GeV}.
\label{eq:cuts1}
\end{align}
The $x$-, $z$- and $M_h$-dependent asymmetries are plotted in the left, central, and right panels of Fig.~\ref{fig:asy1}.
We find that the overall tendency of the asymmetry at the EIC is similar to that at COMPASS.
Although the size of the asymmetry is smaller due to the higher-twist nature of the asymmetry, it is still measurable at the kinematics of EIC.

\section{Conclusion}

\label{Sec.conclusion}

In this work, we have studied the single longitudinal-spin asymmetry with a $\sin\phi_R$ modulation of dihadron production in SIDIS.
We not only considered the contributions from the coupling of the twist-3 distributions $h_L$ and the DiFF $H_{1,ot}^{\sphericalangle}$, but also took into account the coupling of the helicity distribution $g_1$ and the twist-3 DiFF $\widetilde{G}^{\sphericalangle}$.
Using the same spectator model which has been adopted to calculate DiFFs $D_{1,oo}$ and $H_{1,ot}^{\sphericalangle}$, we calculated the twist-3 T-odd DiFF $\widetilde{G}^{\sphericalangle}_{ot}$ by considering the gluon rescattering effect.
We found that the contribution to $\widetilde{G}^{\sphericalangle}_{ot}$ comes from the interference of the $s$ and $p$ waves.
We also test the reliability of the spectator model by comparing the model results of the dihadron fragmentation function $H_{1}^{\sphericalangle}$ with the existed parametrization as well as the model results of the twist-3 distribution function $h_L$ with the Wandzura-Wilczek approximation.
Using the numerical results of the DiFFs, we estimated the $\sin\phi_R$ asymmetry and compared it with the COMPASS measurement.
Our calculation shows that the $h_L\,H_{1,ot}^{\sphericalangle}$ term dominates in the most of the kinematical region.
However, the inclusion of the $g_1\,G_{ot}^{\sphericalangle}$ contribution yields a better description of the COMPASS data, especially in the large $M_h$ region.
In addition, we also made a prediction on the $\sin\phi_R$ asymmetry in SIDIS at the typical kinematics of a future EIC.
Our study shows that the twist-3 DiFF $\widetilde{G}^{\sphericalangle}$ should be considered in phenomenological analysis in order to provide a better understanding of the $\sin\phi_R$ asymmetry in dihadron production in SIDIS.
\\
\\
{\bf Acknowledgements} This work is partially supported by the NSFC (China) grant 11575043, by the Fundamental Research Funds for the Central Universities of China. X. W. is supported by the China Postdoctoral Science Foundation under Grant No. 2018M640680 and the NSFC (China) grant 11847217. Y. Y is supported by the Scientific Research Foundation of Graduate School of Southeast University (Grant No.YBJJ1770) and by the Postgraduate Research \& Practice Innovation Program of Jiangsu Province (Grants No. KYCX17\_0043).

\end{document}